\documentclass[12pt]{article}

\usepackage{epsfig}
\usepackage{graphicx}

\def\be{\begin{equation}}
\def\ee{\end{equation}}
\def\beq{\begin{eqnarray}}
\def\eeq{\end{eqnarray}}
\def\a{\alpha}
\def\b{\beta}
\def\d{\delta}
\def\e{\epsilon}
\def\l{\lambda}

\def\r{\rho}
\def\t{\tilde}
\def\vp{\varphi}

\def\F{{\cal F}}
\def\N{{\cal N}}
\def\O{{\cal O}}

\def\({\left (}
\def\){\right )}
\def\[{\left [}
\def\[{\right ]}

\def\V{{\cal V}}

\begin{document}

\begin{titlepage}
\bigskip
\rightline{}
\rightline{hep-th/0503071}
\bigskip\bigskip\bigskip\bigskip
\centerline
{\Large \bf {Holographic Description of AdS Cosmologies}}
\bigskip\bigskip
\bigskip\bigskip

\centerline{\large Thomas 
Hertog\footnote{Hertog@vulcan.physics.ucsb.edu} and 
Gary T. Horowitz\footnote{gary@physics.ucsb.edu}}
\bigskip\bigskip
\centerline{\em Department of Physics, UCSB, Santa Barbara, CA 93106}
\bigskip\bigskip

\begin{abstract}

To gain insight in the quantum nature of the big bang, we study the dual field theory description 
of asymptotically anti-de Sitter solutions of supergravity that have cosmological singularities. 
The dual theories do not appear to have a stable ground state. One regularization of the theory 
causes the cosmological singularities in the bulk to turn into giant black holes with scalar hair. We 
interpret these hairy black holes in the dual field theory and use them to compute a finite temperature 
effective potential. In our study of the field theory evolution, we find no evidence for a ``bounce" 
from a big crunch to a big bang. Instead, it appears that the big bang is a rare fluctuation from a 
generic equilibrium quantum gravity state.

\end{abstract}

\end{titlepage}

\baselineskip=18pt

\setcounter{equation}{0}
\section{Introduction}

One of the main goals of quantum gravity is to provide a better understanding of the big bang singularity in cosmology. 
This is essential for cosmology to be a truly predictive science, that explains how the distinctive features of the 
universe emerged from the early quantum gravitational phase and why they are what they are. A long-standing issue is 
whether cosmological singularities represent a true beginning or end of evolution. More generally, one would like to 
understand how semiclassical spacetime and our usual notion of time arise from the past singularity. Various suggestions 
have been made for how this quantum gravity transition happens. These include the no boundary wave function 
\cite{Hartle83} that describes creation ex nihilo, and the chaotic initial conditions proposed in \cite{Damour00}.
Alternatively, it is possible that evolution essentially continues through the singularity, with an immediate transition 
from a big crunch to a big bang \cite{Gasperini93,Khoury02}.

Since our usual notions of space and time are likely to break down near cosmological singularities,
a more promising approach to study the problem of initial conditions in cosmology is to find a dual description 
in terms of more fundamental variables. In string theory we do not yet have a dual description of real 
cosmologies, but we do have the celebrated AdS/CFT correspondence \cite{Maldacena98} which provides a 
non-perturbative definition of string theory on asymptotically anti-de Sitter (AdS) spacetimes 
in terms of a conformal field theory (CFT). We have recently constructed examples of solutions in 
$\N=8$, $D=4$ supergravity where smooth, asymptotically AdS initial data emerge from a big bang in the past
and evolve to a big crunch singularity in the future \cite{Hertog04b}. The AdS/CFT duality should 
provide a precise framework in which the quantum nature of the cosmological singularities can be understood\footnote{A different approach to holographic cosmology has been discussed in \cite{Banks04}.}.

With standard AdS-invariant  boundary conditions in the bulk, the dual CFT is the usual $2+1$ theory 
on a stack of M2-branes. To construct the AdS cosmologies,  we have generalized the boundary conditions 
on one of the tachyonic scalars in the theory (while preserving the AdS symmetries), which corresponds 
to modifying the CFT by a triple trace operator. In this paper we study the dual field theory to gain 
insight in the quantum nature of the cosmological singularities. 

This field theory appears to have a potential unbounded from below, which makes it difficult to analyze. 
However the fact that one can map the problem of cosmological singularities in quantum gravity to a 
problem in an ordinary nongravitational field theory appears to be a significant advance. We use a 
variety of approaches to study this dual  theory. One regularization of the unbounded potential turns the 
cosmological singularity into a large black hole with scalar hair\footnote{The hair comes from one of the 
tachyonic scalars in the theory. It has been shown \cite{Sudarsky02} there are no hairy AdS black holes where 
the scalar field asymptotically tends to the true minimum of its potential. The first examples of AdS black 
holes with (non-tachyonic) scalar hair were given in \cite{Torii01}.}. We interpret these hairy black holes in the dual field 
theory and use them to compute a finite temperature effective potential.

While a complete quantum understanding of cosmological singularities is still not available, we are led to a 
picture of the big bang as a rare fluctuation in a typical quantum gravity state in which all the Planck scale 
degrees of freedom are excited. This could help explain the origin of the second law of thermodynamics. 
Penrose \cite{Penrose} has long argued that past singularities such as the big bang must be very different from 
future singularities like a big crunch, since the former must correspond to a state of very low entropy. We will indeed find such an asymmetry between past and future singularities.  Whether the entropy is low enough to explain observations
will require a more detailed cosmological model and in particular, a better understanding how semiclassical
spacetime emerges from a generic quantum gravity state.

A brief outline of this paper is as follows. In the next section we review the construction of asymptotically 
anti de Sitter initial data which evolves to a big crunch. In section 3, we discuss the dual field theory 
evolution using several different approximations. A regularization of the field theory is introduced in 
section 4 and the connection with hairy black holes is explored. This leads to the picture of the big bang as a rare fluctuation. The final section contains some directions 
for further work. In the course of our discussion, we introduce several functions of the form $\b(\a)$. 
To avoid confusion, we list all of them in an Appendix, together with their definition.

\setcounter{equation}{0}
\section{Anti-de Sitter Cosmology}

\subsection{Setup}

We consider the low energy limit of M theory with $AdS_4 \times S^7$ 
boundary conditions. The massless sector of the compactification of $D=11$ 
supergravity on $S^7$ is ${\cal N}=8$ gauged supergravity in four dimensions 
\cite{deWit82}. The bosonic part of this theory involves the graviton, 28 gauge 
bosons in the adjoint of $SO(8)$, and 70 real scalars, and admits $AdS_4$ as a 
vacuum solution. It is possible to consistently truncate this theory to include only 
gravity and a single scalar with action \cite{Duff99}
\be\label{4-action}
S=\int d^4x\sqrt{-g}\left[\frac{1}{2}R
-\frac{1}{2}(\nabla\phi)^2 +2+\cosh(\sqrt{2}\phi) \right]
\ee
where we have set $8\pi G=1$ and chosen the gauge coupling so that the AdS radius is one.
The potential has a maximum at $\phi =0$ corresponding to an $AdS_4$ solution 
with unit radius. It is unbounded from below, but small fluctuations have
$m^2 = -2$, which is above the Breitenlohner-Freedman bound $m_{BF}^2 = -9/4$ 
\cite{Breitenlohner82}, so with the usual boundary conditions $AdS_4$ is stable
\cite{Abbott82,Gibbons83}.

\subsection{Boundary Conditions}

We will work in global coordinates in which the $AdS_4$ metric takes the form
\be \label{adsmetric}
ds^2_0 = \bar g_{\mu \nu} dx^{\mu} dx^{\nu}=
-(1+r^2 )dt^2 + {dr^2\over 1+r^2} + r^2 d\Omega_{2}
\ee
In all asymptotically AdS solutions, the scalar $\phi$ decays at large radius as
\be\label{hair4d}
\phi(r)=\frac{\alpha}{r}+\frac{\beta}{r^2}
\ee
where  $\a$ and $\b$ can
depend on the other coordinates. The standard boundary conditions
correspond to either $\a=0$ or $\b=0$ \cite{Breitenlohner82,Klebanov99}.
It was shown in \cite{Hertog04} that $\b_k = -k\a^2$ (with $k$ an arbitrary constant) 
is another possible boundary condition that preserves all the asymptotic AdS 
symmetries. One can consider even more general boundary conditions $\b = \b(\a)$. 
Although these will generically break some of the asymptotic AdS symmetries, 
they are invariant under global time translations. Hence there is still a conserved 
total energy, as we now show.

As discussed in \cite{Hertog04}, the usual definition of energy in AdS diverges 
whenever $\a\ne 0$. This is because the backreaction of the scalar field causes 
certain metric components to fall off slower than usual. In particular, one has
\cite{Hertog04}
\beq 
\label{4-grr}
g_{rr}=\frac{1}{r^2}-\frac{(1+\alpha^2/2)}{r^4}+
O(1/r^5) & \quad g_{tt}=-r^2 -1+O(1/r) \nonumber\\
g_{tr}=O(1/r^2) \qquad \qquad \qquad & \ \  \ \ \ g_{ab}= \bar g_{ab} +O(1/r) 
\nonumber\\
g_{ra} = O(1/r^2) \qquad \qquad \qquad & g_{ta}=O(1/r) \ \ 
\eeq

The expression for the conserved mass depends on the asymptotic behavior 
of the fields and is defined as follows. Let $\xi^\mu$ be a timelike vector which 
asymptotically approaches a (global) time translation in AdS. The Hamiltonian takes 
the form
\be
H = \int_\Sigma \xi^\mu  C_\mu + {\rm surface \ terms}
\ee
where $\Sigma$ is a spacelike surface, $C_\mu$ are the usual
constraints, and the surface terms should be chosen so that the
variation of the Hamiltonian is well defined. The variation of the
usual gravitational surface term is given by
\beq\label{gravch}
\delta Q_{G}[\xi]&=&\frac{1}{2}\oint dS_i
\bar G^{ijkl}(\xi^\perp \bar{D}_j \delta h_{kl}-\delta h_{kl}\bar{D}_j\xi^\perp)
\eeq
where $G^{ijkl}={1 \over 2} g^{1/2} (g^{ik}g^{jl}+g^{il}g^{jk}-2g^{ij}g^{kl})$,
$h_{ij}=g_{ij}-\bar{g}_{ij}$ is the deviation from the spatial metric 
$\bar{g}_{ij}$ of pure AdS,  $\bar{D}_i$ denotes covariant differentiation 
with respect to $\bar{g}_{ij}$ and $\xi^\perp = \xi \cdot n$ with $n$ the
unit normal to $\Sigma$.  Since our scalar field is falling off more slowly than 
usual if $\a \ne 0$, there is an additional scalar contribution to the surface terms.
Its variation is simply
\be\label{deltaQ}
\delta Q_\phi[\xi] =-\oint \xi^\perp \delta \phi D_i \phi dS^i \ee
Using the asymptotic behavior (\ref{hair4d}) this becomes
\be
\delta Q_\phi[\xi] = r\oint(\a\delta \a)d\Omega 
+ \oint [\delta (\a\b) + \b\delta \a]d\Omega
\ee
Since there is a term proportional to the radius of the sphere, this scalar surface
term diverges. However, this divergence is exactly canceled by the
divergence of the usual gravitational surface term (\ref{gravch}).
The total charge can therefore be integrated, yielding
\be\label{scalarst}
Q [\xi] = Q_{G}[\xi]+ r\oint {\a^2\over 2}d\Omega  + 
\oint [\a\b + W(\a)]d\Omega \ee
where we have defined
\be W(\a) = \int_0^\a \b(\tilde \a)d\tilde \a \ee
In addition to canceling the divergence in (\ref{scalarst}), the
gravitational surface term contributes a finite amount $M_0$. For
the spherically symmetric solutions we consider below, this is
just the coefficient of the $1/r^5$ term in $g_{rr}$. Since $\a$
and $\b$ are now independent of angles, the total mass becomes \cite{Hertog05b}
\be\label{mass}
M= 4\pi(M_0 + \a\b + W)
\ee

\subsection{Big Bang/Big Crunch AdS Cosmologies}

We now review the big bang/big crunch AdS cosmologies of \cite{Hertog04b} that are 
solutions of (\ref{4-action}) with boundary conditions 
\be\label{bdycond}
\b_k=-k\a^2
\ee
on the scalar field. One first finds an $O(4)$-invariant Euclidean instanton solution of the form
\be \label{inst}
ds^2 = {d\r^2\over b^2(\r)} +\r^2 d\Omega_3
\ee
and $\phi=\phi(\r)$. The field equations determine $b$ in terms of $\phi$
\be \label{inst1}
b^2(\r) = { 2V \r^2 -6\over \r^2 \phi'^2 -6}
\ee
and the scalar field $\phi$ itself obeys
\be\label{inst2}
b^2 \phi'' + \left( {3 b^2 \over \r} +bb' \right) \phi' -V_{,\phi} =0
\ee
where prime denotes $\partial_{\rho}$.

Regularity at the origin requires $\phi'(0) =0$.
Thus the instanton solutions can be labeled by the value of $\phi$ at the origin. 
For each $\phi(0)$, one can integrate (\ref{inst2}) and get an instanton.
Asymptotically one finds $\phi(\r)= \alpha/\r +\beta/\r^2$, where $\a$ and $\b$ are now constants. Hence for each $\phi(0)$
one obtains a point in the $(\a,\b)$ plane. Repeating for all $\phi(0)$ yields a curve 
$\b_i(\a)$ where the subscript indicates this is associated with instantons. This curve is 
plotted in Fig 1. (Since the potential $V(\phi)$ is even, it suffices to consider positive 
$\phi(0)$ which corresponds to positive $\a$.) 

The slice through the instanton obtained by restricting to the equator of the $S^3$ 
defines time symmetric initial data for a Lorentzian solution. The Euclidean radial 
distance $\r$ simply becomes the radial distance $r$ in this initial data. So
given a choice of boundary condition $\b(\a)$, one can obtain suitable initial data
by first selecting the instanton corresponding to a point where the curve $\b_i(\a) $ intersects $\b(\a)$, and then  taking a slice through this instanton.

\begin{figure}[htb]
\begin{picture}(0,0)
\put(48,250){$\beta_i$} 
\put(396,244){$\alpha$}
\end{picture}
\mbox{\epsfxsize=14cm \epsfysize=9cm \epsffile{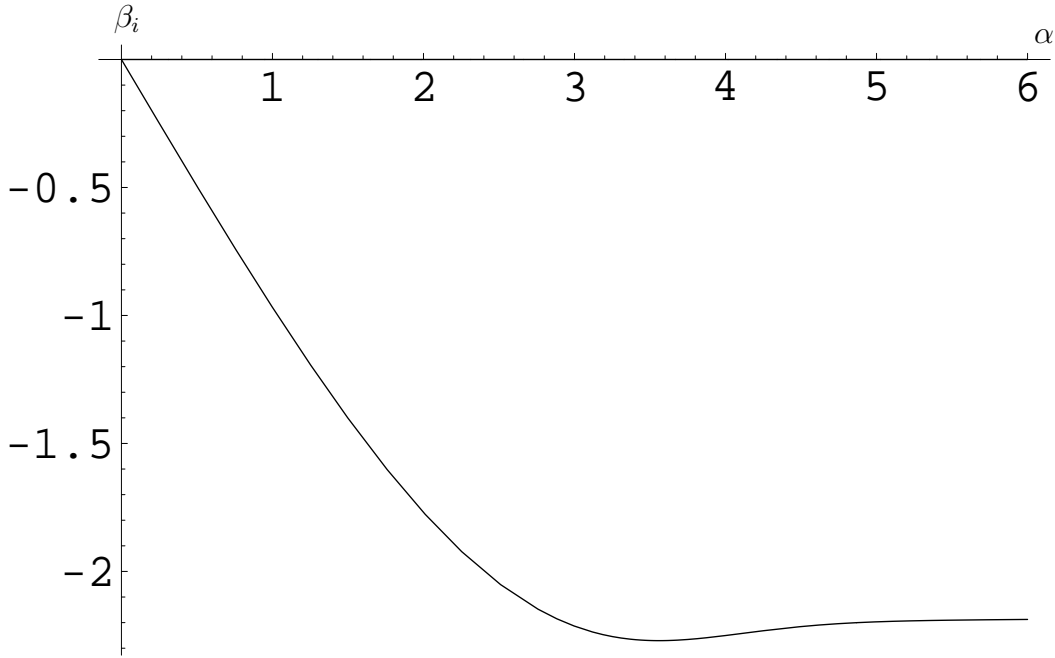}}
\caption{The function $\beta_i$  obtained from the instantons.} 
\label{1}
\end{figure}

Clearly all AdS-invariant boundary conditions (\ref{bdycond}) admit precisely one instanton 
solution. Furthermore, the mass (\ref{mass}) of the resulting initial data is given by
\be \label{mass4dhair}
M=4\pi \left( M_0-\frac{4}{3}k\alpha^3 \right).
\ee
From the asymptotic form of (\ref{inst1}) it follows that $M_0= 4k\alpha^3/3$. Thus this 
initial data corresponds to  a zero mass solution, consistent with its interpretation as the solution 
$AdS_4$ decays into.\footnote{One can also show that this instanton has finite action 
\cite{Hertog04b} which is large for small $k$ and small for large $k$.}

With $\b_k=-k\a^2$ boundary conditions, the evolution of the initial data defined by the instanton
geometry is simply obtained by 
analytic continuation \cite{Coleman:1980aw}. The origin of the Euclidean instanton
becomes the lightcone of the Lorentzian solution. Outside the lightcone, the solution 
is given by (\ref{inst}) with $d\Omega_3$ replaced by three dimensional de Sitter space. 
The scalar field $\phi$ remains bounded in this region. On the light cone we have 
$\phi=\phi(0)$ and $\partial_{t}\phi=0$ (since $\phi_{,\r}=0$ at the origin in the instanton).
Inside the lightcone, the $SO(3,1)$ symmetry ensures that the solution evolves like an open 
FRW universe,
\be \label{ametric}
ds^2 = -dt^2+ a^2(t) d\sigma_3
\ee
where $d\sigma_3$ is the metric on the three dimensional unit hyperboloid.
Under evolution $\phi$ rolls down the negative potential. This causes the scale factor 
$a(t)$ to vanish in finite time, producing a big crunch singularity. 

To verify that this analytic continuation indeed satisfies our boundary condition, we must do a coordinate transformation in the asymptotic region outside the light cone.
 The relation between the usual static coordinates
(\ref{adsmetric}) for $AdS_4$ and the $SO(3,1)$ invariant coordinates,
\be
ds^2 = {d\r^2\over 1+\r^2} + \r^2 (-d \tau^2 + \cosh^2\tau d\Omega_2)
\ee
is
\be
\r^2 = r^2 \cos^2 t -\sin^2 t
\ee
Hence the asymptotic behavior of $\phi$ in global coordinates is given by
\be \label{asscalar}
\phi(r) = {\tilde \alpha\over r} - {k\tilde\alpha^2\over r^2} +O(r^{-3})
\ee
where $\tilde\alpha= \alpha/\cos t$. This clearly satisfies the boundary condition (\ref{bdycond}), but $\tilde \a$ is now time dependent and  blows up as
$t \rightarrow \pi/2$, when the singularity hits the boundary. 

For the purpose of understanding cosmological singularities in M theory, one can forget 
the origin of this solution as the analytic continuation of an instanton. We have simply found 
an explicit example of asymptotically AdS initial data which evolves to a big crunch. Since the initial data is time symmetric, there is also  a big bang in the past (see Fig 2).

\begin{figure}[htb]
\quad \qquad \qquad \mbox{\epsfxsize=9cm \epsfysize=10cm \epsffile{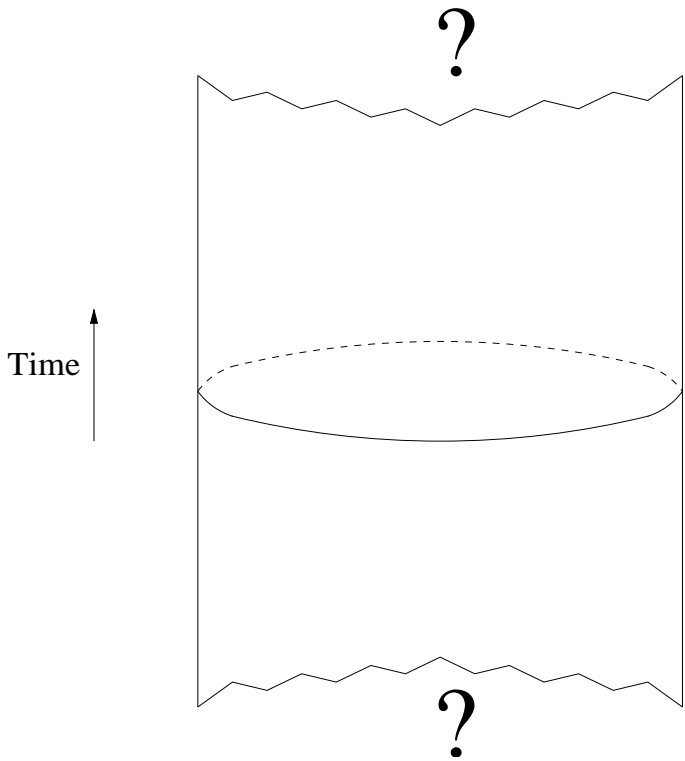}}
\caption{Anti-de Sitter cosmologies.} 
\label{2}
\end{figure}

This theory also has static spherically symmetric solitons \cite{Hertog05b}. Another family of asymptotically AdS cosmologies can be found by starting with the profile 
$\phi_s (r)$ of one of these solitons. Rescaling the soliton configuration $\phi_s$ to 
$\phi_\l(r) = \phi_s(\l r)$ with sufficiently small $\l$ gives negative mass initial data 
for time-dependent solutions \cite{Hertog04b}. Since for small $\l$ we have a large
central region where $\phi$ is essentially constant and away from the maximum of the potential,
it follows that the field must evolve to a spacelike singularity. The 
singularity that develops cannot be hidden behind an event horizon, because all 
spherically symmetric, asymptotically AdS black holes have positive mass \cite{Hertog04}.
Instead, one expects it to continue to spread, cutting off all space. This construction also leads 
to singularities in both the past and future. In addition, there should be solutions with only one singularity, e.g., a big crunch in the future which approach a soliton in the past.

\setcounter{equation}{0}
\section{Dual Field Theory Evolution}

\subsection{Dual CFT}

We now turn to the dual field theory interpretation. M theory on spacetimes which 
asymptotically approach $AdS_4\times S^7$ is dual to the 2+1 conformal field theory (CFT)
describing the low energy excitations of a stack of  $N$ M2-branes. In this correspondence, scalar modes with  $\b=0$ boundary 
conditions correspond to physical states, and adding nonzero $\b$ corresponds to modifying the CFT action. Our bulk scalar $\phi$ is dual to a dimension one operator $\O$. One way of 
obtaining this CFT is by starting with the field theory on a stack of D2-branes and
taking the infrared limit. In that description \cite{Aharony:1998rm},
\be \label{operator}
\O = {1 \over N} Tr T_{ij}\vp^i\vp^j 
\ee
where $T_{ij}$ is symmetric and traceless and $\vp^i$ are the adjoint scalars.

Note that physical states are associated with modes with the slower fall-off. This is possible since the mass is tachyonic and close to the Breitenlohner-Freedman bound. The field theory dual to the ``standard" quantization, where physical states are described by
modes with $\phi =\beta/r^2$ asymptotically, can be obtained by adding the double trace term 
${f\over 2}\int \O^2$ to the action \cite{Witten02,Gubser:2002vv}. This is a relevant 
perturbation and the infrared limit is another CFT in which $\O$ has dimension two. 
As described in \cite{Hertog04}, the AdS invariant boundary conditions $\b_k=-k\a^2$
correspond instead to adding a triple trace term to the action
\be\label{Ocubed}
S = S_0 - { k \over 3} \int \O^3
\ee
The extra term has dimension three, and hence is marginal and preserves conformal invariance, 
at least to leading order in $1/N$. In general, imposing nontrivial boundary conditions $\b (\a)$ in the bulk
corresponds to adding a multi-trace interaction $\int W(\O)$ to the CFT action, 
such that after formally replacing $\O$ by its expectation value $\a$ one has \cite{Witten02,Berkooz} 
\be \label{multitrace}
\b = {\d W \over \d \a}
\ee

\subsection{Semiclassical Field Theory Evolution}

We have seen that the coefficient 
$\tilde\alpha$ of the bulk solution (\ref{asscalar}) diverges as $t\rightarrow \pi/2$, 
when the big crunch singularity hits the boundary. Since the 
coefficient of $1/r$ is interpreted as the expectation value of $\O$
in the dual CFT, this shows that to leading order in $1/N$, $\langle \O\rangle$ diverges in finite time.

A qualitative explanation for this field theory behavior is the following.
The term we have added to the action is not positive definite.  Since the
energy associated with the asymptotic time translation in the bulk can
be negative \cite{Hertog04} (and is in fact unbounded from below\footnote{One can obtain solutions with arbitrarily negative energy by  taking initial data for a static soliton  and rescaling the radial variable. The rescaling  does not change our boundary condition (\ref{bdycond}).}),  the dual field theory should also admit negative energy states and have a spectrum unbounded from below. This shows
 that the usual vacuum must be unstable, and that there are 
(nongravitational) instantons which describe its  decay.
 After the
tunneling, the field rolls down the potential and becomes infinite in finite time. 

A semiclassical analysis supports this reasoning.
If we neglect for a moment the nonabelian structure and identify $\O$ with $\vp^2$,
we are led to consider a single scalar field theory with standard kinetic term and potential,
\be \label{pot}
V= {1 \over 8}  \vp^2 - { k \over 3}\vp^6
\ee
where the quadratic term corresponds to the conformal coupling of $\vp$ to the curvature of $S^2$,\footnote{Since we have set the AdS radius equal to one, the dual field theory lives on $S^2\times R$ where the sphere also has unit radius.} 
and the second term represents the second term in (\ref{Ocubed}). Although this is clearly 
a huge simplification of the field theory, at the classical level it captures the bulk 
behavior in a surprisingly quantitative way. In particular, it admits the following exact
homogeneous classical solution,
\be
\vp (t) = {C \over \cos^{1/2} t}
\ee
where $C=\(3/8k\)^{1/4}$. This solution has zero energy since the field starts at rest where the potential vanishes. Hence it is analogous to the solution obtained by analytically continuing the instanton. Since $\vp^2$ is identified with 
$\tilde\alpha$ on the bulk side, the time dependence of this solution agrees with that
predicted from supergravity, including the fact that the field diverges at $t \rightarrow \pi/2$. 
Furthermore, from Fig 1 it follows that for large $\a$, i.e. large $\phi(0)$, we have
$\a \sim k^{-1/2}$, since $\b$ tends to a constant. Remarkably, this scaling is also 
reproduced by the field theory solution above. 

In this model field theory, the usual vacuum at $\vp =0$ is perturbatively stable but nonperturbatively unstable.
 There are 
(nongravitational) instantons which describe the semiclassical decay of the usual
vacuum. For small $k$, the potential barrier is large, and the
instanton action is large. So tunneling is suppressed. For large $k$,
the barrier is small and tunneling is not suppressed. (This agrees with the action of the gravitational instantons in the bulk.)  After the
tunneling, the field rolls down the potential and becomes infinite in finite time.
So a semiclassical analysis suggests that the CFT does not have well defined evolution for all time.

\subsection{Quantum Mechanics}

We have seen that evolution ends in finite time in the semiclassical description of a 
simplified version of the dual field theory. This agrees with the supergravity result. 
This conclusion changes dramatically, however, if one considers the quantum mechanics of the
potential (\ref{pot}). That is, we again concentrate on the homogeneous mode 
$\vp (t)=x(t)$ only, but now treat it quantum mechanically. The quantum mechanics of unbounded
potentials of this type is well understood and discussed in detail in \cite{Reed75,Carreau90}.

Like the classical trajectories, a wave packet with an energy distribution peaked at some 
value $E$ that moves in a potential of the form $-kx^6$ (with $k >0$) will reach infinity in 
finite time. Thus a packet can `disappear' and probability is apparently lost. But in quantum 
mechanics this problem can be dealt with by constructing a self-adjoint extension of the 
Hamiltonian 
$H = -(1/2) (d^2/dx^2) +V(x)$, which is done by carefully specifying its domain. 
Once a domain is chosen, the Hamiltonian is self-adjoint and unitary time evolution is 
guaranteed.
 
One can construct these domains by finding the corresponding set of eigenfunctions. 
At large $x$ the WKB approximation is accurate and the wave functions for each energy $E>0$ are
\be
\chi^{\pm}_{E} (x) = \(2E+2k x^6\)^{-1/4} e_{\ }^{\pm i\int_0^x \sqrt{2E+2k y^6}dy}
\ee
The Hamiltonian is not Hermitian on a domain containing all wave functions of this form.
For instance, if $\phi_1 = \chi_{E}^{+}$ and $\phi_2 = \chi_{E'}^{+}$ or $\chi_{E'}^{-}$ then
$\(H \phi_1, \phi_2 \) \neq \(\phi_1, H \phi_2 \)$. It can be made Hermitian, 
however, by
selecting a particular linear combination of $\chi_{E}^{+}$ or $\chi_{E}^{-}$. 
Let (again at large $x$)
\be
\psi_{E}^{\gamma} =  \(2E+2k x^6\)^{-1/4} \cos \(\int_0^x \sqrt{2E+2k y^6}dy 
                     +\omega^{\gamma}_{E}\)
\ee
where
\be
\omega^{\gamma}_{E} = \gamma -\int_0^{\infty}\( \sqrt{2E+2k y^6} - \sqrt{2k y^6}\)dy
\ee
and $\gamma$ is an arbitrary phase. This choice of $\omega^{\gamma}_{E}$ makes $\psi_{E}^{\gamma}$
approach a real, energy-independent function $\bar \psi^{\gamma}$ at large $x$. The complete set 
$\psi_{E}^{\gamma}$, for fixed $\gamma$, can be used to define the domain of $H$.
The Hamiltonian defined on this domain is Hermitian, since all wave functions in the domain 
approach a (complex) constant times the same real function of $x$. Notice, however, that since 
$\gamma$ is arbitrary, there is a one-parameter family of self-adjoint Hamiltonians, each of which 
results in a different unitary time evolution.
Which self-adjoint extension is chosen by string theory (if any) is an interesting open question.
Our point here is only to show that the quantum mechanical description indicates that the big crunch
is not an endpoint of evolution.

The fact that $H$ is self-adjoint means that probability does not leave the system. Yet the
center of a wave packet follows essentially the classical trajectory and still reaches infinity 
in finite time. What happens \cite{Carreau90} is that a right-moving wave packet bounces off infinity
and reappears as a left-moving wave packet. Hence the quantum mechanics of the potential (\ref{pot}) in the 
dual field description of our AdS cosmologies indicates that the big crunch is not an endpoint of 
evolution. Furthermore, it shows that for exactly homogeneous initial data there is a bounce
through the singularity, as envisioned in the pre-big bang \cite{Gasperini93} and cyclic universe 
\cite{Khoury02} models.
 
There is another approach to applying quantum mechanics to potentials unbounded from below based on a $PT$ 
symmetry \cite{Bender04}. However this approach is motivated by analytic continuation from the harmonic 
oscillator and results in a positive spectrum. In contrast, for all self adjoint extensions described above, 
there is an infinite (discrete) set of negative energies.  Since the bulk theory is known to have negative energy
solutions, the approach using self adjoint extensions seems more appropriate.

\subsection{Full Quantum Field Theory}

We have seen that a quantum mechanical analysis of the homogeneous mode of the dual field indicates that
evolution is unitary and that there is an immediate transition from a big crunch to a big bang. 
It is natural to ask if one should expect this conclusion to hold also in the full field theory. 
It is not known if self-adjoint extensions can be constructed in field theories with
potentials of this form. However, even if one can define a unitary evolution,  
the full field theory evolution is likely to be very different, because our discussion of 
the quantum mechanics obviously neglects the possibility of particle creation.

It is well known that a scalar field that oscillates near the minimum of its effective potential
rapidly converts its energy into particles that are produced during these oscillations.
It was recently found, however, that in many theories where the scalar field rolls down from the top of 
its effective potential towards the true minimum, particles are produced in great numbers while
the field is rolling down. This phenomenon is called tachyonic preheating \cite{Felder01,Felder01b}. 
It happens essentially because the effective negative mass term in the potential causes long wavelength 
quantum fluctuations to grow exponentially.

Tachyonic preheating is so efficient that in many theories most of the initial potential energy density 
is converted into the energy of scalar particles well before the field reaches the true minimum. 
Thus a prolonged stage of oscillations of the homogeneous component of
the scalar field around the true minimum of the potential does not exist in spontaneous symmetry breaking.

Tachyonic preheating occurs, therefore, also in the dual field theory description of our AdS cosmologies,
where the supergravity initial data correspond to a homogeneous field theory configuration high up the
potential. This means that even if the field theory has well-defined evolution for all time, it will in 
general not be dual to a bounce through the singularity. Indeed this would require the miraculous 
conversion of all the energy back into the homogeneous mode! 

As an aside, we note that there is a long history of studying cosmological singularities via a ``minisuperspace 
approximation", in which one first restricts to homogeneous cosmologies and then quantizes the remaining finite 
number of degrees of freedom. It is clear from the above discussion that this approach misses a key physical 
effect\footnote{See \cite{Kuchar} for an earlier objection to minisuperspace.}. The possibility of exciting all the inhomogeneous modes can dramatically change the evolution near a 
singularity. It is amusing to recall that before the discovery of the singularity theorems, it was widely 
believed that (classical) singularities would only arise in highly symmetric collapse, and generic collapse 
would be nonsingular. We now see that the situation in quantum gravity is almost the opposite: a strictly 
homogeneous collapse results in a simple bounce, while a generic collapse does not.

\setcounter{equation}{0}
\section{From Black Holes to Big Crunch}

\subsection{Regularized Field Theory}

The main difficulty in analyzing our dual field theory is clearly that the potential is unbounded from below. 
If there was a stable ground state, a homogeneous field rolling down the steep potential would evolve to a 
thermal state about the global minimum of the potential. One can regulate our potential and produce a stable 
ground state by, e.g.,  
 adding to the CFT lagrangian
 \be
 W(\O) = -{k\over 3} \int \O^3 +   {\epsilon\over 4}\int \O^4
 \ee
 Neglecting again the 
nonabelian structure this gives the following potential
\be \label{pot2}
V= {1 \over 8} \vp^2 - { k \over 3}\vp^6 + {\epsilon \over 4} \vp^8
\ee
Provided $\epsilon$ is sufficiently small, this has a local minimum at $\vp =0$ and 
a global minimum at $\vp^2 \sim k/\epsilon$.

The operator $\O^4$ is not renormalizable, but to leading order in $1/N$ this does not cause any difficulty.  Its only effect in the bulk is to  modify the scalar boundary condition which according to
(\ref{multitrace}) becomes
\be\label{modbc}
\b_{k,\e}=-k\a^2+\epsilon \a^3
\ee
This modification will  slightly change  our initial data and can affect its evolution. However, its 
effect on evolution can only be appreciable in the ``corners" of Fig 2 where the singularity hits 
the boundary at infinity. This is because the change in the boundary
conditions is negligible until $\a$ becomes large, and causality then restricts its effect to these 
regions. In particular,  since the evolution of the $\epsilon=0$ data has trapped surfaces, a singularity 
will still form in the central region.  However it is now possible that the singularity is enclosed 
inside a large black hole and does not extend out to infinity. We now investigate this possibility.

\subsection{Black Holes with Scalar Hair}

Appropriate initial data for the modified boundary conditions (\ref{modbc}) is obtained by taking an 
instanton with $\b_i(\a)=\b_{k,\e}(\a)$
and slicing it through the center. From Fig 1, it is clear that there are two possible instantons 
satisfying this condition.  We are interested in the one with
$\b \approx -k\a^2$ so the initial data is close to the one we studied in section 2.
The evolution of this initial data cannot be obtained by analytic continuation of the instanton since 
that solution will not satisfy the new boundary conditions. A full description of the evolution probably 
requires numerical relativity. However, we can at least ask if there is a static black hole with the right 
mass to form from our initial data. The instanton field equation (\ref{inst1}) always yields $M_0 =  -4\a\b/3$.  
So from (\ref{mass}), the mass of our new initial data is slightly negative\footnote{The theory with modified boundary 
conditions has negative mass solutions, but it can be shown there is a lower bound on the mass 
and hence a stable ground state \cite{Hertog05c}, as suggested by the dual field theory potential (\ref{pot2}).}
\be
M= 4\pi\left [-{4\over 3} \a\b_{k,\e} + \a\b_{k,\e} - {1\over 3} k\a^3 + {1\over 4} \epsilon \a^4\right] 
= -{\pi\over 3} \epsilon \a^4
\ee
Since the energy is conserved during evolution, a black hole can form only if there exist
negative mass black holes. These black holes will necessarily have scalar hair, so we have to study hairy black 
holes with boundary conditions (\ref{modbc}).

It was shown in \cite{Hertog04} that the theory (\ref{4-action}) with AdS-invariant boundary conditions
(\ref{bdycond}) admits black hole solutions with scalar hair\footnote{See \cite{Martinez04,Radu05} for other examples
of black hole solutions with tachyonic scalar hair.}. For a given choice of $k \neq 0$, there 
is a one-parameter 
family of hairy black holes. The solutions can be characterized by their conserved mass
(\ref{mass4dhair}), which uniquely determines the horizon radius $R_e$ as well as the value 
$\phi_e$ of the scalar field at the horizon of this class of solutions. The hairy black holes 
provide an example of black hole
non-uniqueness, however, since Schwarschild-AdS is also a solution  for all boundary conditions. We will 
discuss below how the dual field theory description resolves this non-uniqueness.

The hairy black holes all have positive mass (\ref{mass4dhair}) in this theory \cite{Hertog04}, so the zero 
mass initial data defined by the instantons cannot evolve to a black hole. Indeed we have seen that with 
AdS-invariant boundary conditions, initial data of this type produce a big crunch. This is not
necessarily the case, however, if one evolves with the modified boundary 
conditions (\ref{modbc}). We now show that with boundary conditions $\b_{k,\e}$,  the bulk theory admits a 
second branch of regular black hole solutions with scalar hair, which includes negative mass black holes. 

\begin{figure}[htb]
\begin{picture}(0,0)
\put(43,251){$\beta_{R_e}$}
\put(407,245){$\alpha$}
\end{picture}
\mbox{\epsfxsize=14cm \epsfysize=9cm \epsffile{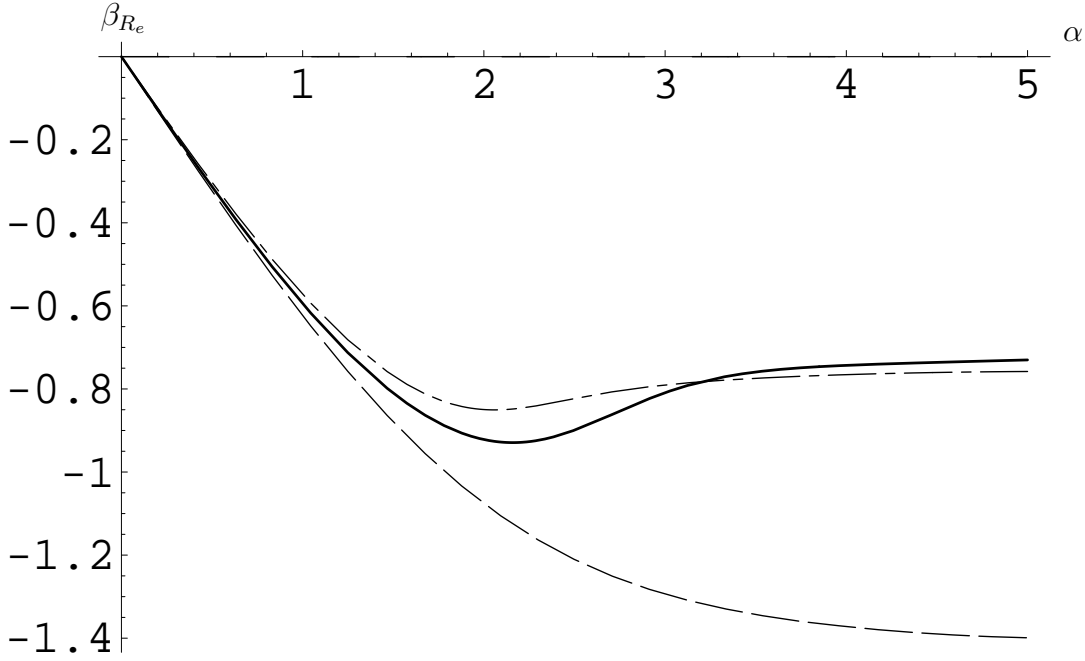}}
\caption{The functions $\beta(\alpha)$ obtained from the solitons 
and from hairy black holes of two different sizes. The full line shows the soliton curve 
$\b_{s}(\a)$, the dot-dashed line shows the $\b_{R_e}(\a)$ curve for $R_e=.2$ black holes and 
the dashed line is the $R_e=1$ curve.}
\label{3}
\end{figure}

We find the hairy black holes by numerically integrating the field equations for static,
spherically symmetric solutions. Writing the metric as
\be
ds^2=-h(r)e^{-2\delta(r)}dt^2+h^{-1}(r)dr^2+r^2d\Omega_2
\ee
the Einstein equations read
\be\label{hairy14d}
h\phi_{,rr}+\left(\frac{2h}{r}+\frac{r}{2}\phi_{,r}^2h+h_{,r}
\right)\phi_{,r}   =  V_{,\phi}
\ee
\be\label{hairy24d}
1-h-rh_{,r}-\frac{r^2}{2}\phi_{,r}^2h =  r^2V(\phi)
\ee
\be
\delta_{,r} = -{1 \over 2}r \phi_{,r}^2
\ee
We integrate the field equations outward from the horizon.
Regularity at the event horizon $R_e$ requires
\be \label{horcon4d}
\phi_{,r}(R_{e}) = {R_{e}V_{,\phi}(\phi_e) \over 1-R_{e}^2V(\phi_{e})}
\ee

The scalar field asymptotically behaves as (\ref{hair4d}), so we obtain a point in the $(\a,\b)$ plane
for each combination $(R_e,\phi_e)$.  Repeating for all $\phi_e$ gives a curve $\b_{R_e}( \a)$.  
In Fig 3 we show this curve for hairy black holes of two different sizes. We also show 
the curve obtained in a similar way for regular solitons, which were discussed in \cite{Hertog05b}. 
As one increases $R_e$, the curve decreases faster and reaches larger (negative) values of $\beta$.
Given a choice of boundary conditions $\b (\a)$, the allowed black hole solutions are simply given 
by the points where the black hole curves intersect the boundary condition curve: $\b_{R_e}(\a) = \b (\a)$. 

\begin{figure}[htb]
\begin{picture}(0,0)
\put(27,254){$\beta_{k,\e}$}
\put(405,191){$\alpha$}
\end{picture}
\mbox{\epsfxsize=14cm \epsfysize=9cm \epsffile{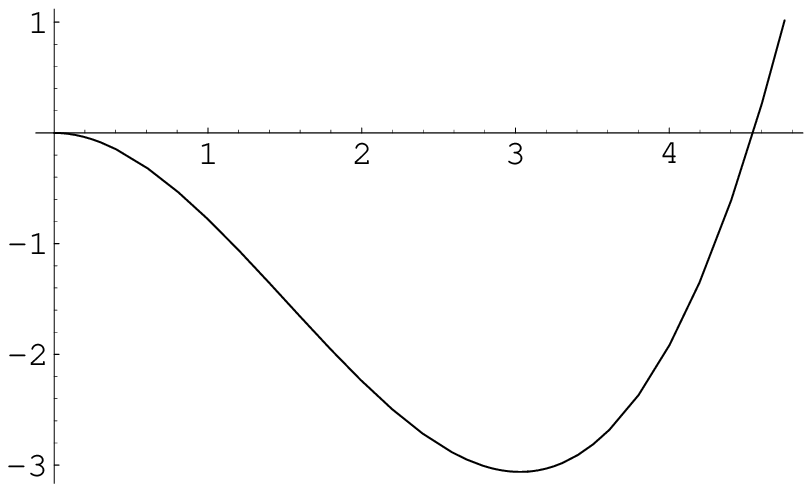}}
\caption{The boundary condition curve $\b_{k,\e}(\a) = -k\a^2 +\e\a^3$ with $k=1$ and 
$\e=.22$.}
\label{4}
\end{figure}

From the black hole curves shown in Fig 3, it follows immediately that with AdS-invariant boundary
conditions $\b_k =-k\a^2$, there is precisely one hairy black hole solution for each 
radius $R_e$, as was discussed in \cite{Hertog04}.
By contrast, the curve $\b_{k,\e}$ shown in Fig 4, which is defined by the modified boundary conditions 
(\ref{modbc}), intersects the black hole curves twice for small $R_e$. On the other hand, it is clear there 
are no large hairy black holes that obey the $\b_{k,\e}$ boundary conditions. The size of the largest hairy 
black hole solution increases for decreasing $\epsilon$.

\begin{figure}[htb]
\begin{picture}(0,0)
\put(32,249){$M/4\pi$}
\put(410,50){$R_e$}
\end{picture}
\mbox{\epsfxsize=14cm \epsfysize=9cm \epsffile{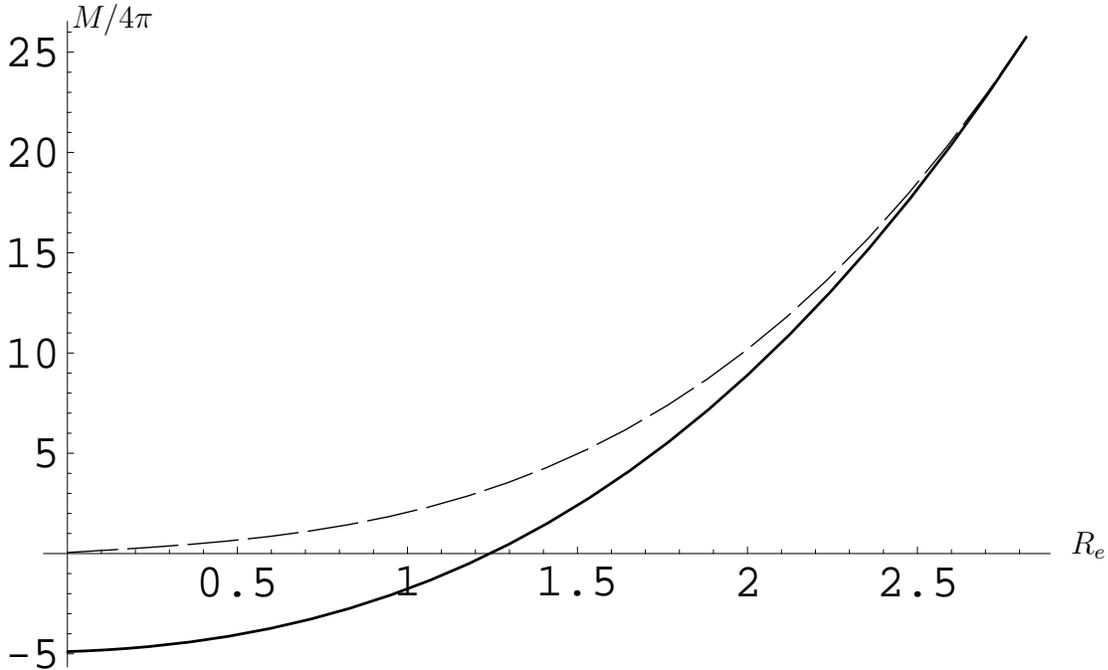}}
\caption{The mass of the hairy black holes that obey the boundary conditions $\b_{k,\e}(\a)$ with $k=1$ and $\e =.22$.. 
The full line gives the masses of the second branch of solutions, which are associated with the second 
intersection point of the curves $\b_{k,\e}(\a)$ and $\b_{R_e}(\a)$, and hence have more hair. 
This branch disappears in the limit $\e \rightarrow 0$.}
\label{5}
\end{figure}

The small $\a$ (and hence small $\phi_e$) black holes form one branch of solutions, which also exist 
in the theory with $\b_k =-k\a^2$ boundary conditions. 
Their mass is positive, as shown in Fig 5 with $k=1$ 
and $\epsilon=.22$, and in fact they are always more massive than a Schwarschild-AdS black hole of the same 
size \cite{Hertog04}. On the other hand, the solutions with the larger $\a$, which are associated with the 
second intersection point, have much smaller mass. In fact,  provided $\epsilon$ is sufficiently small, 
some hairy black holes on this second branch have negative mass, as we illustrate in Fig 5. 
There is thus a black hole with scalar hair on the second branch with the same mass as our initial data.  
This is the natural endstate of the evolution of the initial data defined by the instanton\footnote{The instanton
initial data that obey $\b_{k,\e}(\a)$ boundary conditions with $k=1$ and $\e=.22$ have mass $M/4\pi = -.057$.}.

Support for this comes from the fact that as $\e$ decreases, the second branch of black holes moves down to 
lower mass. Hence, the size of the black hole with mass equal to our initial data increases. In the limit 
$\e \rightarrow 0$, the black hole becomes infinitely large. Furthermore, $\phi_e \rightarrow \phi(0)$, where 
$\phi(0)$ is the value of the scalar field on the lightcone that emanates from the origin of the instanton 
initial data. Therefore, the bulk evolution with boundary conditions defined by the modified field theory with a 
stable ground state does not describe the formation of a cosmological singularity, but rather a large black hole 
with scalar hair. As one removes the regulator $\epsilon$, the minimum of the potential 
approaches minus infinity and the size of the black hole diverges.

\subsection{ Dual Field Theory Description of Hairy Black Holes }

The hairy black holes found above have a natural description in terms of the dual field theory which we now 
describe. It was shown in \cite{Hertog05b} how to compute the effective potential for the vacuum expectation 
value of the dual operator ${\cal O}$ in presence of arbitrary deformations $W({\cal O})$. If $S_0$ is the 
action of the usual 2+1 CFT that is dual to M theory with $\b=0$ boundary conditions, and
\be\label{deform}
S= S_0 + \int W(\O) 
\ee
then  the expectation values of ${\cal O}$ in different vacua are
obtained by finding nonsingular bulk solitons with boundary conditions $\b=W'$ \cite{Witten02,Berkooz}. 
Given a soliton with $\b=W'$, 
one has $\langle \O\rangle = \a$. To find the effective potential, one starts with the curve
 $\b_{s}(\a)$ obtained from the regular soliton solutions and shown in Fig 3. 
 One then defines a function
\be
W_0(\a) = - \int_{0}^{\a} \b_{s}(\t \a)d\t\a
\ee
and sets 
\be
\V=W_0+W
\ee
It follows immediately that the extrema of $\V$ are in one-to-one correspondence with solitons that obey 
the boundary conditions $\b=W'(\a)$.  So the location of the extrema yield  $\langle \O\rangle$. Furthermore, one can show that the value of $\V$ at the extremum gives the energy of the corresponding soliton and hence also the energy of the dual  field theory state \cite{Hertog05b}.  Therefore one can interpret $\V$ as the effective potential for $\langle \O\rangle$.
The fact that the function $W$ does not receive any corrections in the effective potential
is  reminiscent of a nonrenormalization theorem. 

The left panel of Fig 6 shows the effective potential in the deformed field theory that we discussed earlier,
\be\label{dform}
 W (\O) = -{1 \over 3}  \O^3 + {\e\over 4}  \O^4
\ee
with $\e=.22$
which corresponds to the modified  boundary conditions $\b_{k,\e}$ in the bulk.
Note that $\V$ has three extrema: a local minimum at $\a=0$ and two extrema with $\a\ne 0$.  So this field theory has three different vacua  and one can consider excitations about each. The usual Schwarzschild-AdS black holes correspond to a typical excitation of mass $M$ about the $\a=0$ vacuum\footnote{This applies to most black holes, but not the very small ones.  For small excitation energy, the typical state corresponds to a thermal gas surrounding the soliton. }. The top branch of the hairy black holes in Fig 5 corresponds to excitations about the local maximum of $\V$, and the bottom branch  corresponds to excitations about the global minimum. This interpretation is motivated by several facts. First,   for a given size black hole, the top branch is clearly more massive than the bottom. Second, in the limit $\e\rightarrow 0$ where one removes the global minimum, the bottom branch of hairy black holes is absent. The top branch is essentially unchanged (although it now continues for arbitrarily large $R_e$). This is consistent with the fact that the local maximum of the potential is essentially unchanged in this limit. Finally, and most importantly, the upper branch of black holes is unstable.  We have shown that there is an unstable spherically symmetric scalar perturbation.  This instability is directly analogous to the instability found for hairy black holes with $\e=0$ boundary conditions \cite{Hertog05}.  In constrast, this unstable mode is not present for the hairy black holes in the lower branch, and we believe they are stable. This agrees with the stability of the corresponding field theory vacua.

Further information about the black hole states in the dual field theory can be obtained by considering the expectation value of $\O$ in the black hole state. We can define an effective potential whose extrema are precisely these expectation values by generalizing our earlier discussion of the vacuum expectation values. Recall that by considering all hairy black holes with horizon radius $R_e$, one obtains a curve $\b_{R_e}(\a)$ (see Fig 3).
 For each radius  we  define a function
\be
W_{R_e}(\a) =   R_e (1+R_e^2) - \int_{0}^{\a} \b_{R_e}(\t\a) d\t\a
\ee
This function is universal, in the sense that it is independent of the choice of boundary conditions. For any  $W(\a)$,   the function 
\be\label{Repot}
{\cal V}_{R_e}(\a) = W_{R_e}(\a)+W(\a), 
\ee
clearly has extrema precisely when there are black holes of size $R_e$ that obey the boundary conditions $\b=W'(\a)$. So the location of the extrema give the expectation value of $\O$ in the black hole state.
In addition, we now show that the the value of ${\cal V}_{R_e}$ at each extrema gives the mass of the corresponding black hole and hence the energy of the field theory state. Suppose we choose our boundary condition to be
$\b = \b_{R_e} (\a)$. In this case, all hairy black holes of size $R_e$ are allowed by the boundary conditions\footnote{In fact, there are no black holes of any other size in this theory. }.
Furthermore, since the area is fixed, the first law of black hole mechanics implies $dM=TdS =0$. In other words,  all black holes have the same mass in this theory, which  equals the mass of the Schwarschild-AdS black hole with radius $R_e$ (which is the $\a=\b=0$ point on the curve) . 
Setting  $M= 4\pi R_e (1+R_e^2)$  and using (\ref{mass}) it  follows that
\be
W_{R_e} = M_0 +\a \b 
\ee
Hence for general boundary conditions $\b=W'(\a)$ one obtains
\be \label{m2}
M = 4\pi (W_{R_e} +W ) = \oint \V_{R_e} d\Omega
\ee
where we have used the fact that $\b =\b_{R_e}(\a)$ for all black hole solutions.
Thus for general boundary conditions $\b(\a)$ the mass of the black holes (including Schwarschild-AdS)
with these boundary conditions is given by the value of $\V_{R_e}$ at the corresponding extrema. Hence  $\V_{R_e}$ is an effective potential for the expectation value of $\O$ in the black hole state.

\begin{figure}[htb]
\begin{picture}(0,0)
\put(18,168){$\V_{0}$}
\put(202,156){$\alpha$}
\put(251,141){$\V_{1}$}
\put(425,65){$\alpha$}
\end{picture}
\mbox{\epsfxsize=7cm \epsfysize=6cm \epsffile{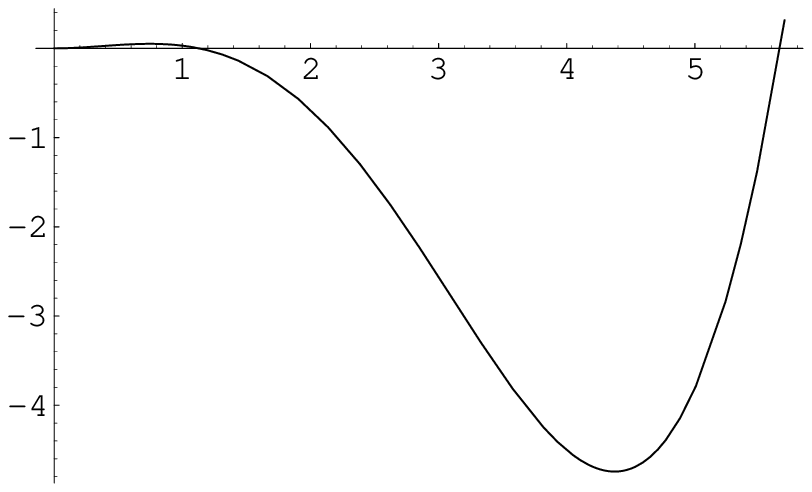} \qquad 
\epsfxsize=7cm \epsfysize=5cm \epsffile{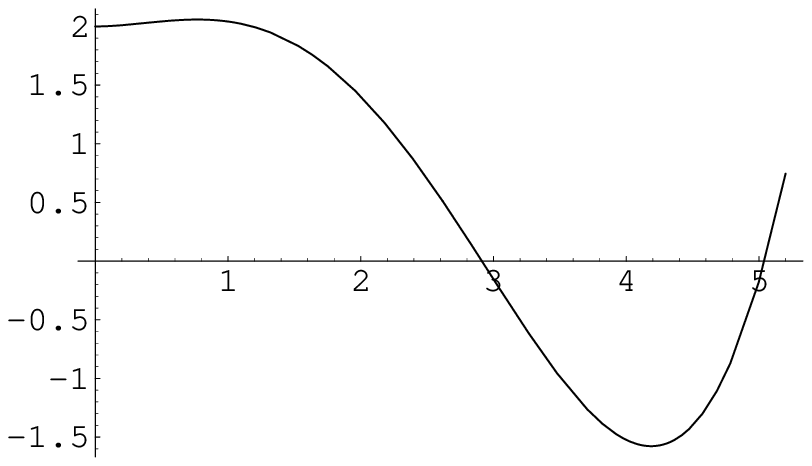}}
\caption{The left panel shows the effective potential $\V_0$ for the vacuum expectation values $\langle \O \rangle$ 
in the field theory with deformation $W =-{1 \over 3} \a^3 +.055 \a^4$.
The right panel shows the effective potential $\V_1$ for $\langle \O \rangle$ in the same theory, but with the 
expectation value taken in finite energy states dual to different black holes of size $R_e=1$.}
\label{6}
\end{figure}

In the right panel of Fig 6 we plot the effective potential $\V_1$, constructed from the $R_e=1$ curve, in the deformed field theory (\ref{dform}).  One sees that like the vacuum
effective potential, $\V_1$ has a local minimum at $\a=0$ and two additional extrema at $\a = \langle \O \rangle \neq 0$.  The value $\V_1 (0)$ equals the mass of Schwarschild-AdS with $R_e=1$, while 
$\V_1$ at its local maximum/global minimum gives the masses of respectively the
$R_e=1$ hairy black hole on the upper/lower branch in Fig 5. Note that 
 the distance 
$\Delta \a$ between both $\a \neq 0$ extrema as well as the difference $\Delta \V$ between their energies are smaller in $\V_1$ compared to $\V_0$. These differences further decrease for increasing mass.
For sufficiently massive hairy black holes the extrema of the corresponding effective potential eventually merge and then disappear, which in the bulk corresponds to the fact that there is a maximum size hairy black hole (where the two branches in Fig 5 meet).

\subsection{Finite Temperature Field Theory}

In the context of the AdS/CFT correspondence, rather than working with states of fixed energy as above,  
black holes are often discussed in the context of a canonical ensemble, in which they are dual to thermal 
states in the field theory. We now show how to compute 
a finite temperature effective potential in the dual field theory.

\begin{figure}[htb]
\begin{picture}(0,0)
\put(45,247){$\beta_{T}$}
\put(400,240){$\alpha$}
\end{picture}
\mbox{\epsfxsize=14cm \epsfysize=9cm \epsffile{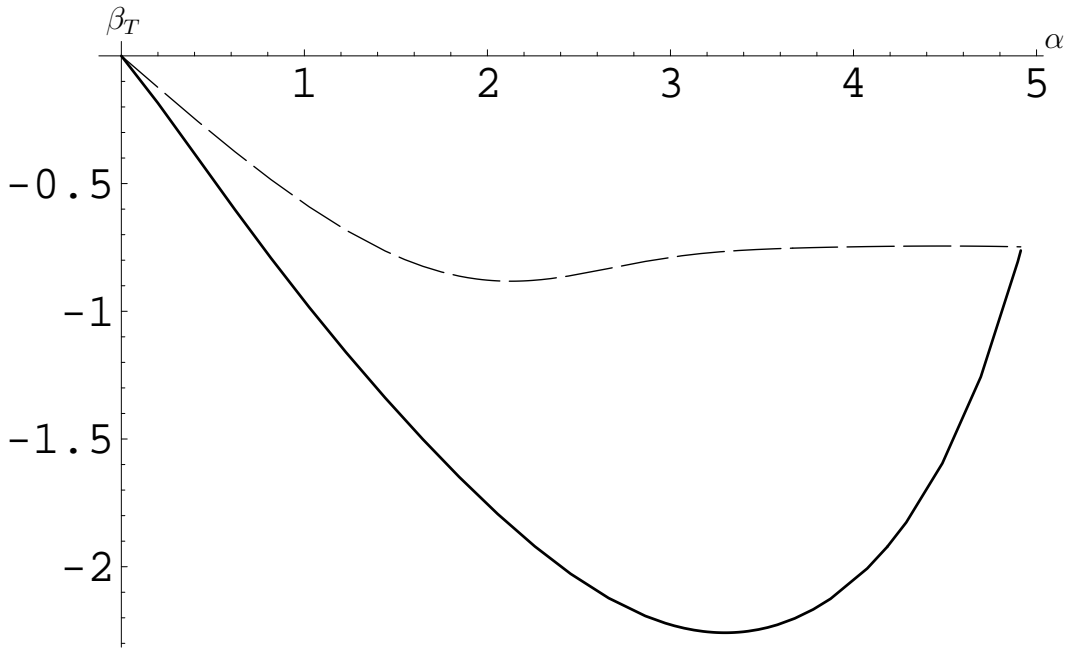}}
\caption{The constant temperature curve $\b_{T} (\a)$ for $4\pi T =7$.}
\label{7}
\end{figure}

The temperature of a hairy black hole is given by
\be\label{temp}
4\pi T = { 1-V(\phi_e)R_e^2 \over R_e} e^{\delta_{\infty}}
\ee
where 
\be
\delta_{\infty} = - {1 \over 2} \int_{R_e}^{\infty} r \phi_{,r}^2 dr
\ee
One sees that, like Schwarschild-AdS black holes, hairy black holes only contribute to the thermodynamic 
ensemble at sufficiently high temperatures, $2\pi T \geq \sqrt{-V(\phi_e)}  e^{\delta_{\infty}}$.  
By adjusting $R_e$ and $\phi_e$ so that the temperature (\ref{temp}) is held fixed, one finds  a one-parameter 
family of 
hairy black holes with the same temperature $T$. From the asymptotic value of the scalar field, we obtain 
a curve  $\b_{T}(\a)$  shown in  Fig 7  for  $4\pi T=7$. This curve consists of two branches that are 
smoothly connected at the maximum value of $\a$. However, the upper/lower branch tend to a {\it different} 
Schwarschild-AdS solution at the origin of the 
$(\a,\b)$-plane, namely the usual small/large black hole with $4\pi T=7$. In general, the upper branch, 
which tends to the soliton curve for $ T \rightarrow \infty$, represents the smaller black holes associated 
with a given temperature, while the lower branch contains the larger black holes (we note, however, that 
the radius $R_e$ is not constant along the curves $\b_{T}(\a)$). Thermal states in the field theory are 
dual to the larger black hole, so we focus on the lower branch of this curve. As $T$ increases, this lower 
branch moves down and the maximum value of $\a$ increases, so the area enclosed by the two constant temperature curves 
increases too.

The lower branch $\b_{T} (\a)$ can be used to define a function
\be\label{Wtemp}
W_T (\a) = { \bar F_T\over 4\pi} - \int_{0}^{\a} \b_{T} (\t\a) d\t\a,
\ee
where the constant $\bar F_T $ is the free energy of the larger 
Schwarschild-AdS black hole of temperature $T$.  As before,  for any $W(\a)$, the extrema of
\be
\F_{T}(\a) = W_T(\a) +W(\a)
\ee
correspond precisely to hairy black holes of temperature $T$ that obey the boundary condition $\b=W'(\a)$. 
In addition, we now show that the value of $\F_{T}$ at each extremum gives the free energy of the 
corresponding hairy black hole and hence also the dual field theory configuration. Suppose the boundary 
conditions were $\b = \b_{T}(\a)$. Then all the black holes along the lower branch constant $T$ curve are 
allowed. The first law $dM=TdS$ now implies that the free energy $M-TS$ is constant along this 
curve.\footnote{Since the upper and lower curves in Fig 7 join smoothly at a maximum $\a$, one might 
expect the free energy to be constant along the entire curve. However this would contradict the fact 
that the two Schwarzschild-AdS black holes of the same temperature have different free energies. The 
resolution is that one must fix the boundary conditions.  The free energy is constant only along a 
continuous family of black holes with the same temperature in the same theory.  Since the boundary 
conditions must be a single valued function of $\a$, the upper and lower branch cannot both be present 
in one theory.}  The value of this constant is fixed by the $\a\rightarrow 0$ limit to be the 
Schwarzschild-AdS value so 
\be
W_T (\a) +{TS(\a)\over 4\pi}= M_0 +\a \b 
\ee
Hence for general boundary conditions $\b =W'(\a)$ we have
\be\label{free}
F_{T} = 4\pi (W_T +W) =\oint \F_{T} d\Omega
\ee

\begin{figure}[htb]
\begin{picture}(0,0)
\put(52,248){$\F_T$}
\put(401,60){$\alpha$}
\end{picture}
\mbox{\epsfxsize=14cm \epsfysize=9cm \epsffile{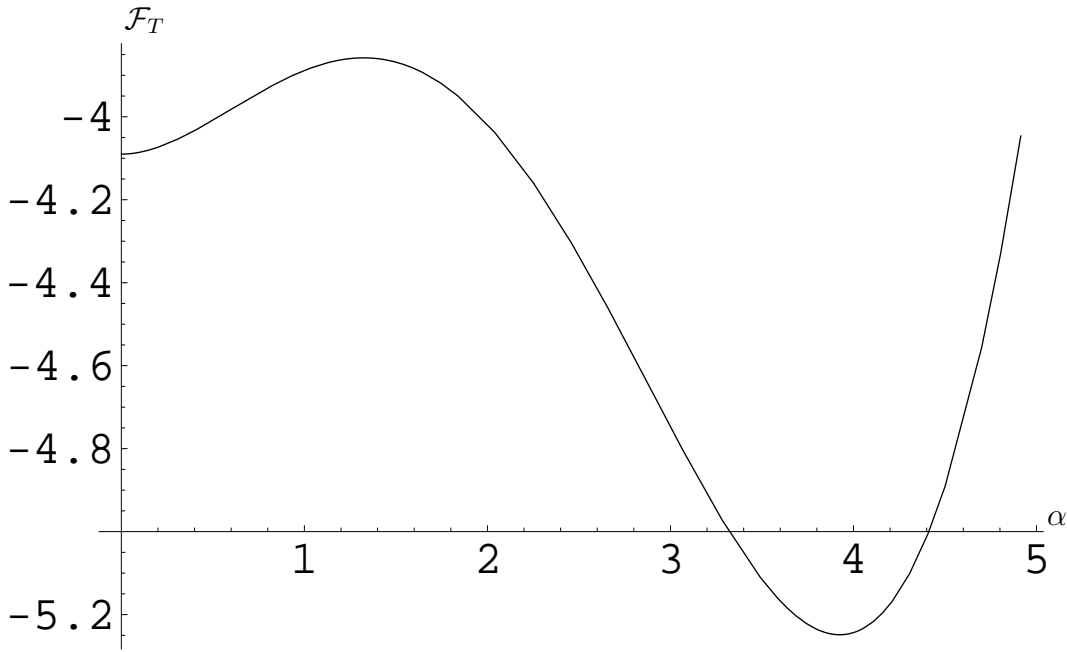}}
\caption{The effective potential for the free energy $\F_{T} = W_{T}-{1 \over 3} \a^3 +.055 \a^4$ 
at temperature $4\pi T=7$, constructed from the lower branch of the $\b_{T} (\a)$ curve.}
\label{8}
\end{figure}

Therefore in the dual field theory one can interpret  $\F_{T}$ as the finite temperature
effective potential. We illustrate this in Fig 8, where we plot the effective potential $\F_{T}$ 
for $4\pi T=7$ in the field theory  dual to $\b_{k,\e}$ boundary conditions.
We see that the hairy black holes that are thermal states in the global minimum are thermodynamically 
favoured.

\subsection{The Big Bang as a Rare Fluctuation}

We now return to the dual description of our AdS cosmologies. We have seen that a modification 
of the  boundary conditions in the bulk, corresponding to the regularization of the dual field theory, 
turns the big crunch into a giant black hole with scalar hair. Since the mass of this black 
hole is negative and much larger black holes exist with these same boundary conditions (see Fig 5) 
we must ask if the black hole we form corresponds to the larger or smaller one of the given temperature. 
It is easy to see that for small $\e$ we always form the larger black hole. This is because in 
this regime, $\phi_e $ remains bounded and $R_e$ is very large, so that the temperature of our black hole is very high.
 The other black hole of the same temperature would instead have to be very small. Hence the 
black hole we form indeed corresponds to a thermal state in the field theory. As expected, the 
formation of the large hairy black hole in the bulk  corresponds on the field theory side  to the zero 
mode rolling down the potential, exciting all the inhomogeneous modes, and eventually producing a 
thermal state in the new global minimum that arises from the regularization. 

Notice that the evolution in the bulk is nearly independent of $\e$ for 
while. In particular, the light cone of the origin of the (time symmetric) initial data, expands out to 
large radius in all cases. If $\e$ is nonzero, it eventually stops expanding and becomes the event horizon, 
while if $\e=0$, it continues to expand and reaches infinity at the same time as the big crunch. This shows 
that the approach to the big crunch is identical to the formation of a large black hole.  In the dual field 
theory, this is the statement that when the field rolls down the potential, it does not know if there will 
be a global minimum or not. 

If one introduces a large radius cut-off in the bulk, one cannot tell the difference between a very large black hole and a big crunch. But this IR cut-off in the bulk corresponds to a UV cut-off in the dual field theory. This suggests that the evolution to a big crunch can also be viewed as evolving to an equilibrium  state in the dual theory. One can view this equilibrium state in the bulk as having all the Planck scale degrees of freedom excited.  It would not correspond to any semiclassical spacetime.

For the field to roll back up the potential would require an exceedingly rare fluctuation, which 
converts all the energy of the approximately thermal state back into the homogeneous mode. In the 
bulk this would correspond to the time reversed evolution in which there is a big bang singularity  in our past.  This would mean that the big bang in our  AdS cosmologies should 
be viewed as a rare fluctuation from a generic equilibrium state in quantum gravity.  One can imagine that the boundary theory spends a long time in this equilibrium state, which does not describe any semiclassical spacetime.  A semiclassical spacetime only arises after a rare fluctuation which, in the dual 
field theory, corresponds to most of the energy going back into the zero mode. 

This picture leads to a natural asymmetry between past and future singularities: the evolution 
from a past singularity requires a rare fluctuation which, in our context, causes the zero mode to shoot 
up the potential. In contrast, the approach to a big crunch is the generic evolution corresponding to the
field rolling down the potential. 

It is natural to speculate that this asymmetry is more general. If so, this could help explain the origin 
of the second law of thermodynamics in realistic cosmologies. Penrose has stressed that our universe started 
in an extremely low entropy state\footnote{His estimate for the maximum possible entropy comes from putting all the matter in the observable universe into a large black hole. This is similar to the large black hole in our regularized theory.}, and suggested that this was a result of a fundamental difference between 
past and future singularities \cite{Penrose}. The picture we are led to here is similar. However, while Penrose 
has argued that one needs a time asymmetry in the laws of nature to explain this difference, this does not seem 
to be present in AdS/CFT.  

Clearly, the fact that a rare fluctuation is required for a semiclassical spacetime is not sufficient to explain 
the second law, since the universe today is semiclassical and has much larger entropy than the early universe. 
To make further progress one needs a better understanding of the quantum gravitational transition that describes 
the emergence of semiclassical spacetime from the generic quantum gravity state we envision. This would allow
one to compute the relative probabilities of different universes.

If a semiclassical spacetime is the result of a rare fluctuation, then the absolute probability for any 
feature of the observed universe would be exceedingly small. Of more interest would be conditional 
probabilities, in which one assumes some gross feature of the universe is present and then asks about the 
probability for other features \cite{Hawking02}.  One could, for instance, ask whether an expanding 
universe like ours is likely to have an early period of inflation. At present the literature contains 
conflicting statements on this issue.
 
Dyson et. al. \cite{Dyson02} have argued that if the big bang is a rare fluctuation then it 
would be more likely for a universe like ours to depend on ``statistical miracles", instead of evolving in 
a way that can be understood by usual physical reasoning. However, their starting point was a low temperature 
thermal state in de Sitter space. De Sitter holography (the idea that one causal patch includes all the degrees 
of freedom) then implied that a fluctuation that produced the big bang and an inflationary universe must include 
all the degrees of freedom, which made such fluctuation extremely unlikely. 

On the other hand, Albrecht et al. \cite{Albrecht04} have performed an alternative calculation, in which they take the 
same starting point but calculate the relative probabilities from more traditional semiclassical tunneling rates 
(instead of invoking a principle of causal patch physics). In contrast with \cite{Dyson02}, they find that inflation is 
strongly favored over other paths to our observed universe.

Finally, Hartle and Hawking have put forward a definite proposal for the wave function of the universe 
\cite{Hartle83}. This provides the most concrete framework to date to compute probabilities for different
semiclassical spacetimes. One finds again that the observed universe is more likely to have an early inflationary
phase than to arise from a fluctuation directly in its present state \cite{Hartle04}. Furthermore, one can also compare 
the relative probabilities of different inflationary histories. In theories where the inflaton potential has a maximum, 
for example, one finds a universe like ours is most likely to emerge in a de Sitter state via the Hawking-Moss instanton 
with a homogeneous field at the maximum \cite{Hawking02}. 

The Hartle-Hawking wave function is peaked around semiclassical geometries. It is therefore likely to differ from the 
equilibrium quantum gravity state\footnote{The generic equilibrium quantum gravity state we envision can be viewed as 
a wave function that defines initial condition for cosmology.} we propose, in which the a priori probability for 
semiclassical spacetime would be exceedingly small. However, the predictions of probabilities {\it conditioned} on there 
being semiclassical spacetime may well turn out to be in good agreement\footnote{Possible connections between the 
Hartle-Hawking wave function and string theory were also discussed in \cite{Horowitz04} and \cite{Ooguri05}.}. It would be 
interesting to see if further work on the AdS cosmologies yields a more concrete understanding of the wave function. 
This could potentially place one of the existing proposals for the initial conditions in cosmology on firmer footing. A 
few directions to obtain a more complete understanding of the quantum description of AdS cosmology are outlined in the 
next section.

\setcounter{equation}{0}
\section{Discussion}

We have seen that the dual description of an AdS cosmology involves a field theory with a potential that 
is unbounded from below. We have used various approaches to study its  dynamics.  In a semiclassical 
analysis, the evolution ends in finite time. In a full quantization of the homogeneous mode, evolution 
continues for all time and suggests a bounce. However this is an artefact of throwing away all the 
inhomogeneous degrees of freedom.  If one regulates the potential so that it has a global minimum, one sees all 
the inhomogeneous modes become excited  and the system evolves into a thermal state about the true vacuum. 
However, the corresponding bulk evolution now produces a large black hole with scalar hair rather than a 
cosmological singularity. The cosmological singularity arises as the limit of a specific class of hairy 
black holes as the regulator $\e$ is taken to zero. This suggests that the approach to a big crunch is naturally viewed in 
the dual theory as a similar evolution to an equilibrium state, but one which does not describe a semiclassical spacetime.

We also saw in section 4 how one can use hairy black holes in the bulk to compute various effective potentials 
in the dual field theory. Another application of this discussion  is to designer gravity \cite{Hertog05b}. 
We have recently shown that one can 
``pre-order" solitons in the theory (\ref{4-action}) in the sense that for {\it any} function $\V(\a)$ with $\V(0)=0$, 
there are boundary conditions such that the gravitational theory has solitons precisely at the extrema of 
$\V$ with mass given by the value of $\V$ at the extrema. It is easy to show that one can pre-order hairy 
black holes as well, either in terms of their radius or temperature. For example, suppose one wants to 
specify the mass of hairy black holes of radius $R_e$ in the theory (\ref{4-action}).  Given any function 
$\V_{R_e}(\a)$ with $\V_{R_e}(0)$ equal to the Schwarzschild-AdS mass,  one defines $W(\a)$ via eq. (\ref{Repot}) 
and chooses boundary conditions $\b=W'$. It follows from our earlier discussion that the resulting theory 
has hairy black holes of this radius at each extremum of $\V_{R_e}(\a)$ and the mass of the black hole will 
be given by the value of this function at its extrema.

The dual description of AdS cosmologies suggests that the big bang and the emergence of semiclassical spacetime 
is an exceedingly rare fluctuation from a generic equilibrium state in quantum gravity.
A few possible directions to obtain a more complete quantum description of AdS cosmologies are the following. 
Clearly one needs a better understanding of the equilibrium field theory state when the regulator $\e$ is taken to zero.
Since we are driven to arbitrarily large values of the fields in this limit, one 
might wonder if the AdS/CFT correspondence breaks down, and whether the dual field theory must be extended to include 
more stringy (or M-theory) effects\footnote{We thank D. Gross for suggesting this possibility.}. This is 
suggested by the fact that the original AdS/CFT correspondence arose by taking a low energy limit of the 
excitations of a stack of branes, and we are being driven to consider large excitations. Another interesting 
direction is to study the BKL chaos in AdS cosmology. It has been shown that in the full 11D supergravity evolution, 
the generic approach to a cosmological singularity is chaotic with different spatial points decoupling and undergoing a 
series of Kasner oscillations \cite{Damour00}. It would be interesting to relate this to the dynamics of tachyonic 
preheating in the dual field theory. This relation will have to include the fact that our truncated theory of four 
dimensional gravity coupled to a scalar does not exhibit this chaotic behavior \cite{Andersson,Erickson04}. 
Finally, Banks and Fischler \cite{Banks04} have given a description of the big crunch which has some 
similarities to ours. In particular, they propose the big crunch is a state which does not correspond to a semiclassical 
spacetime, maximizes the entropy, and is related to a conformal symmetry.  
It would be interesting to explore if there is a deeper connection.

\bigskip

\centerline{{\bf Acknowledgments}}
\bigskip

It is a pleasure to thank T. Banks, M. Kleban, J. Maldacena, J. Polchinski, R. Roiban, N. Seiberg,  S. Shenker, and 
E. Witten for discussions. This work was supported in part by NSF grant PHY-0244764.

\appendix

\section{Glossary}

The functions $\b(\a)$ refer to the asymptotic behavior of the bulk scalar field (\ref{hair4d}).
We list here the various functions that are used in the text:
\begin{itemize}
\item{The function $\b_i(\a)$ is defined by the $O(4)$-invariant instantons. It is shown in Fig 1.}
\item{The function $\b_s(\a)$ is defined by the spherical solitons. It is shown in Fig 3.}
\item{The functions $\b_{R_e}(\a)$ are defined by the hairy black holes of size $R_e$. 
A few examples are shown in Fig 3.}
\item{The functions $\b_{T}(\a)$ are defined by the hairy black holes of temperature $T$. 
An example is shown in Fig 7.}
\item{The functions $\b_{k}(\a)$ denote the AdS-invariant boundary conditions $\b_{k}(\a)=-k\a^2$.}
\item{The functions $\b_{k,\e}(\a)$ denote the modified boundary conditions $\b_{k,\e}(\a) = -k\a^2 +\e \a^3$ 
(see Fig 4).}
\item{Without subscript, the function $\b(\a)$ refers to an arbitrary boundary condition.}
\end{itemize}


\begin{thebibliography}{99}


\bibitem{Hartle83}
J. B. Hartle, S. W. Hawking,
``The Wave Function of the Universe,''
Phys. Rev. {\bf D28} (1983) 2960

\bibitem{Damour00}
T. Damour, M. Henneaux,
``Chaos in Superstring Cosmology,''
Phys. Rev. Lett. {\bf 85} (2000) 920, hep-th/0003139;
T.~Damour, M.~Henneaux and H.~Nicolai,
  ``Cosmological billiards,''
  Class.\ Quant.\ Grav.\  {\bf 20} (2003) R145,
hep-th/0212256.

\bibitem{Gasperini93}
M. Gasperini, G. Veneziano, ``Pre-Big Bang in String Cosmology,''
Astropart. Phys. {\bf 1} (1993) 317, hep-th/9211021;
``The pre-big bang scenario in string cosmology,''
Phys.\ Rept.\  {\bf 373} (2003) 1, hep-th/0207130

\bibitem{Khoury02}
J. Khoury, B. A. Ovrut, N. Seiberg, P.J. Steinhardt, N. Turok,
``From Big Crunch to Big Bang,''
Phys. Rev. {\bf D65} (2002) 086007, hep-th/0108187;
 P.J. Steinhardt, N. Turok, ``Cosmic Evolution in a Cyclic Universe,''
Phys. Rev. {\bf D65} (2002) 126003, hep-th/0111098

\bibitem{Maldacena98}
J. M. Maldacena,
``The large N limit of superconformal field theories and supergravity,''
Adv.\ Theor.\ Math.\ Phys.\  {\bf 2} (1998) 231, hep-th/9711200

\bibitem{Hertog04b}
T. Hertog, G. T. Horowitz,
``Towards a Big Crunch Dual,''
JHEP {\bf 0407} (2004) 073, hep-th/0406134

\bibitem{Banks04}
T. Banks, W. Fischler,
``Holographic Cosmology,'' hep-th/0405200;
``The Holographic Approach to Cosmology", hep-th/0412097.

\bibitem{Sudarsky02}
D. Sudarsky, J. A. Gonzales,
``On Black Hole Scalar Hair in Asymptotically Anti de Sitter Spacetimes,''
Phys. Rev. {\bf D67} (2003) 024038, gr-qc/0207069

\bibitem{Torii01}
T. Torii, K. Maeda and M. Narita,
``Scalar hair on the black hole in asymptotically anti-de Sitter spacetime,''
Phys.\ Rev.\ D {\bf 64} (2001) 044007

\bibitem{Penrose}
R. Penrose, in {\it General Relativity: An Einstein Centenary Survey}, eds. S. W. Hawking and W. Israel, Cambridge University Press, 581 (1979);  {\it Proc. 14th Texas Symp. on Relativistic Astrophysics},
ed. E. J. Fergus, New York Academy of Sciences, 249 (1989).

\bibitem{deWit82}
B. de Wit, H. Nicolai,
``$N=8$ Supergravity with Local $SO(8) \times SU(8)$ Invariance,''
Phys. Lett. {\bf 108B} (1982) 285; ``$N=8$ Supergravity,''
Nucl. Phys. {\bf B208} (1982) 323

\bibitem{Duff99}
M. J. Duff, J. T. Liu,
``Anti-de Sitter Black Holes in Gauged N=8 Supergravity,''
Nucl. Phys. {\bf B554} (1999) 237, hep-th/9901149

\bibitem{Breitenlohner82}
P.~Breitenlohner and D.~Z.~Freedman, ``Stability In Gauged Extended 
Supergravity,'' Annals Phys.\  {\bf 144} (1982) 249;
``Positive Energy In Anti-De Sitter Backgrounds And Gauged Extended
Supergravity,'' Phys.\ Lett.\ B {\bf 115} (1982) 197

\bibitem{Abbott82}
L. F. Abbott, S. Deser,
``Stability of Gravity with a Cosmological Constant,''
Nucl. Phys. {\bf B195} (1982) 76

\bibitem{Gibbons83}
G.~W.~Gibbons, C.~M.~Hull and N.~P.~Warner,
``The Stability Of Gauged Supergravity,''
Nucl.\ Phys.\ B {\bf 218} (1983) 173

\bibitem{Klebanov99}
I.R. Klebanov, E. Witten, 
``AdS/CFT Correspondence and Symmetry Breaking,''
Nucl. Phys. {\bf B556} (1999) 89, hep-th/9905104

\bibitem{Hertog04}
T. Hertog, K. Maeda,
``Black Holes with Scalar Hair and Asymptotics in $N=8$ Supergravity,''
JHEP {\bf 0407} (2004) 051, hep-th/0404261

\bibitem{Hertog05b}
T. Hertog, G. T. Horowitz,
``Designer Gravity and Field Theory Effective Potentials,''
hep-th/0412169

\bibitem{Coleman:1980aw}
S.~R.~Coleman and F.~De Luccia,
``Gravitational Effects On And Of Vacuum Decay,''
Phys.\ Rev.\ D {\bf 21} (1980) 3305

\bibitem{Aharony:1998rm}
O.~Aharony, Y.~Oz and Z.~Yin,
``M-theory on AdS(p) x S(11-p) and superconformal field theories,''
Phys.\ Lett.\ B {\bf 430} (1998) 87, hep-th/9803051

\bibitem{Witten02}
E. Witten, ``Multi-Trace Operators, Boundary Conditions, and AdS/CFT 
Correspondence,'' hep-th/0112258

\bibitem{Gubser:2002vv}
S.~S.~Gubser and I.~R.~Klebanov,
``A universal result on central charges in the presence of double-trace
deformations,''
Nucl.\ Phys.\ B {\bf 656} (2003) 23, hep-th/0212138


\bibitem{Berkooz}
  M.~Berkooz, A.~Sever and A.~Shomer,
 ``Double-trace deformations, boundary conditions and spacetime
 singularities,''
  JHEP {\bf 0205} (2002) 034,
  hep-th/0112264;
   A.~Sever and A.~Shomer,
  ``A note on multi-trace deformations and AdS/CFT,''
  JHEP {\bf 0207} (2002) 027,
 hep-th/0203168.


\bibitem{Reed75}
M. Reed, B. Simon,
``Methods of Modern Mathematical Physics II: Fourier Analysis, Self-Adjointness,''
Academic Press, New York (1975)

\bibitem{Carreau90}
M. Carreau, E. Fahri, S. Gutmann, P. F. Mende,
``The Functional Integral for Quantum Systems with Hamiltonians Unbounded from Below,''
Ann. Phys. {\bf 204} (1990) 186

\bibitem{Bender04}
  C.~M.~Bender and S.~Boettcher,
  ``Real Spectra in Non-Hermitian Hamiltonians Having PT Symmetry,''
  Phys.\ Rev.\ Lett.\  {\bf 80} (1998) 5243,
  physics/9712001;
C.~M.~Bender, D.~C.~Brody and H.~F.~Jones,
  ``Complex Extension of Quantum Mechanics,''
  Phys.\ Rev.\ Lett.\  {\bf 89} (2002) 270401
  [Erratum-ibid.\  {\bf 92} (2004) 119902]
  quant-ph/0208076.


\bibitem{Felder01}
G. Felder, J. Garcia-Bellido, P. B. Greene, L. Kofman, A. Linde, I. Tkachev,
``Dynamics of Symmetry Breaking and Tachyonic Preheating,''
Phys. Rev. Lett. {\bf 87} (2001) 011601, hep-ph/0012142

\bibitem{Felder01b}
G. Felder, L. Kofman, A. Linde,
``Tachyonic Instability and Dynamics of Spontaneous Symmetry Breaking,''
Phys. Rev. {\bf D64} (2001) 123517, hep-th/0106179

\bibitem{Kuchar}
  K.~V.~Kuchar and M.~P.~Ryan,
  ``Is Minisuperspace Quantization Valid?: Taub In Mixmaster,''
  Phys.\ Rev.\ D {\bf 40} (1989) 3982.

\bibitem{Martinez04}
C. Martinez, R. Troncoso, J. Zanelli,
``Exact Black Hole Solution with a Minimally Coupled Scalar Field,''
Phys. Rev. {\bf D70} (2004) 084035, hep-th/0406111

\bibitem{Radu05}
E. Radu, D. H. Tchrakian,
``New Hairy Black Hole Solutions with a Dilaton Potential,''
Class. Quant. Grav. {\bf 22} (2005) 879, hep-th/0410154

\bibitem{Hertog05c}
T. Hertog, S. Hollands, D. Marolf, work in progress

\bibitem{Hertog05}
T. Hertog, K. Maeda,
``Stability and Thermodynamics of AdS Black Holes with Scalar Hair,''
Phys. Rev. {\bf D71} (2005) 024001, hep-th/0409314

\bibitem{Hawking02}
S. W. Hawking, T. Hertog,
``Why Does Inflation Start at the Top of the Hill?,''
Phys. Rev. {\bf D66} (2002) 123509, hep-th/0204212

\bibitem{Dyson02}
L. Dyson, M. Kleban, L. Susskind,
``Disturbing Implications of a Cosmological Constant,''
JHEP {\bf 0210} (2002) 011, hep-th/0208013

\bibitem{Albrecht04}
A. Albrecht, L. Sorbo,
``Can the Universe Afford Inflation?,''
Phys. Rev. {\bf D70} (2004) 063528, hep-th/0405270

\bibitem{Hartle04}
J. B. Hartle, private communication

\bibitem{Horowitz04}
G. T. Horowitz, J. Maldacena,
``The Black Hole Final State,''
 JHEP {\bf 0402} (2004) 008, hep-th/0310281

\bibitem{Ooguri05}
H. Ooguri, C. Vafa, E. Verlinde,
``Hartle-Hawking Wave Function for Flux Compactifications,''
hep-th/0502211

\bibitem{Andersson}
  L.~Andersson and A.~D.~Rendall,
  ``Quiescent cosmological singularities,''
  Commun.\ Math.\ Phys.\  {\bf 218} (2001) 479,
  gr-qc/0001047


\bibitem{Erickson04}
J. K. Erickson, D. H. Wesley, P. J. Steinhardt, N. Turok,
``Kasner and Mixmaster Behavior in Universes with Equation of State $w \geq 1$,''
Phys. Rev. {\bf D69} (2004) 063514, hep-th/0312009


  


\end{thebibliography}
\end{document}